\documentstyle[aps,epsfig,twocolumn]{revtex}

\title{Probing Entanglement and Non-locality of Electrons\\ in a Double-Dot
via
Transport and Noise}
\author{Daniel Loss and Eugene V. Sukhorukov}
\address{ Department of Physics and Astronomy, University of Basel,\\
Klingelbergstrasse 82, CH-4056 Basel, Switzerland}

\date{\today}

\begin{document}
\twocolumn[\hsize\textwidth\columnwidth\hsize\csname
@twocolumnfalse\endcsname

\maketitle

\begin{abstract}
Addressing the feasibilty of quantum communication with electrons we
consider
entangled spin states of electrons in a double-dot  which is weakly coupled
to
in--and outgoing leads. We show that the entanglement of two electrons
in the  double-dot  can be detected in mesoscopic
transport and noise measurements. In the Coulomb blockade and cotunneling
regime
the singlet and
triplet states lead to  phase-coherent current and noise contributions of
opposite
signs and to  Aharonov-Bohm and 
Berry phase oscillations in response
to magnetic
fields. These oscillations are a genuine two-particle effect
and provide a direct
measure of  non-locality  in  entangled states. We
show that the ratio of zero-frequency noise
to current (Fano factor) is universal and equal to the electron charge.
\end{abstract}


\vskip2pc]
\narrowtext

Entanglement  and  non-locality of Einstein-Podolsky-Rosen (EPR)
pairs\cite{Einstein} are  remarkable
features of quantum mechanics which give rise to
striking phenomena such as violation of Bell inequalities
and  secure quantum communication\cite{Bennett84}.
Such and related phenomena have been  tested with great
success for photons\cite{Aspect,Zeilinger}, but not yet for {\it massive}
particles such as electrons, particularly  in a solid state
environment. There are two immediate problems. First we
need a scheme by which  entanglement of electrons can be generated in a
controlable manner. In the context
of quantum computing\cite{Steane98} we have recently shown that such a scheme
can be realized in tunnel-coupled quantum dots each of which contains one
single (excess) electron whose spin defines the qubit
\cite{Loss98,Burkard,Imamoglu}.
There are several motivations for such a qubit scheme, most notably
very long spin  decoherence times\cite{Kikkawa97}, and the qubit
defined as electron-spin is mobile and thus is a good candidate for
implementing quantum communication schemes which are
based on EPR pairs\cite{DiVincenzoLossMMM}.  An important
feature of this scheme, moreover,  is  that we have control over the  non-locality
of the
entangled state: electron 1 and 2 are {\it localized} in {\it different} quantum
dots and while being spatially
separated from each other their total spin ground state (a singlet) is
entangled \cite{delocalized}.

The second problem which then immediately arises is: How   can we
probe entanglement?  In the following we 
provide an answer to
this question and show that 
this  property  can
be tested via mesoscopic transport and noise measurements.
The  scheme we propose consists
of two coupled quantum dots---double-dot (DD) for short---which
themselves are weakly coupled  to two leads 1 and 2 (see Fig. 1).  In
contrast to
earlier set-ups involving a DD\cite{blick,Oosterkampmolecule}, we
propose
a scheme where an electron coming from lead 1 (2) has the option to tunnel
into
{\it both} dot 1 and dot 2.
This results in
a closed loop, and applying a magnetic field, an Aharonov-Bohm (AB) phase
$\varphi$
will be accumulated by an electron traversing the DD. In the Coulomb blockade
(CB) regime we find that due to
\begin{figure}
\narrowtext
{\epsfxsize=8.cm
\centerline{\epsfbox{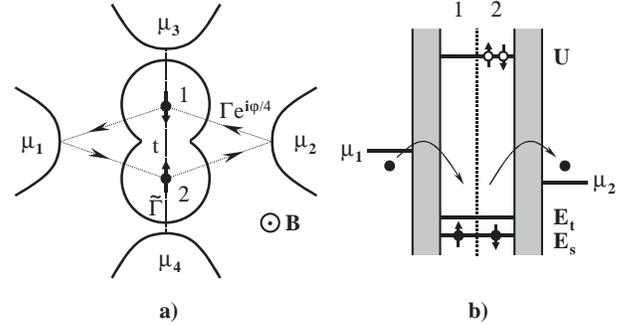}}}
\caption{a) Double-dot (DD) system containing two electrons and
being weakly coupled to metallic leads 1,...,4, each of which is at the
chemical
potential
$\mu_1$,...,$\mu_4$. The tunneling amplitudes between dots and leads are
denoted
by $\Gamma$, ${\tilde \Gamma}$. The tunneling (t) between the dots
results in a singlet ground state\protect\cite{Burkard} with energy $E_s$
and a
triplet state with $E_t$, with $J=E_t-E_s$ describing the effective exchange
coupling between the electron spins (qubits). The closed tunneling path
between dots
and leads 1 and 2 encloses the area $A$. b)
Energy diagram of the DD system described in a)
in the CB regime where  cotunneling is dominant, {\it i.e.},
$U>|\mu_1\pm \mu_2|>J$, with $U$ being the Coulomb
repulsion within a single dot.}
\label{fig1}
\end{figure}
cotunneling\cite{averinazarov} the current and
the noise  depend on the state of the DD:
the AB oscillations for singlet and triplets have opposite sign.
The amplitude of the AB oscillations provides a  measure of
the phase coherence of the entangled state, while the period
--via the enclosed
area--provides a  measure  of the non-locality of the EPR pairs.
The triplets
themselves can  be further distinguished by applying a directionally
inhomogeneous
magnetic field which adds a Berry phase leading to beating.
Finally, we also discuss  finite frequency noise, and show in particular
that
the power spectrum for a DD  contains a
non-vanshing odd-frequency part which is  sensitive to entanglement.

{\it Model System.\/}
The DD system (see Fig. 1) contains $4$ metallic leads which are
in equilibrium with associated reservoirs kept at the chemical
potentials $\mu_i$, $i=1,\ldots ,4$, where the  currents $I_i$ can be measured.
The leads are weakly coupled to the dots
with amplitudes $\Gamma$ and $\tilde{\Gamma}$, and  the leads
$1,2$  are coupled to {\it both} dots and play the role
of probes where the current and noise are measured.
The leads $3$ and $4$ are feeding
electrodes to manipulate
the electron filling in the dots.
The quantum dots contain one (excess) electron each, and
are coupled to each other by the tunneling amplitude $t$, which
leads to a level splitting\cite{Loss98,Burkard} $J=E_t-E_s\sim 4t^2/U$ of
the
single-particle energy levels in the DD,  with $U$ being the
single-dot
Coulomb repulsion energy, and $E_{s/t}$ are the singlet/triplet
energies\cite{footnoteenergy}.
We recall that for two electrons in the DD
(and for weak magnetic fields) the ground state is given by a spin
singlet\cite{Loss98,Burkard}.
For convenience we count the chemical potentials $\mu_i$ from $E_s$.
  The coupling $\tilde{\Gamma}$ to
the feeding leads can be be switched off while probing the DD
with a current.
{}From now on we assume that $\tilde{\Gamma }=0$ unless stated otherwise.

Using a standard tunneling Hamiltonian approach\cite{Mahan},
we write
$H=H_0+V$, where  the first term in $H_0=H_D+H_1+H_2$  describes
the DD and $H_{1,2}$ the leads (assumed to be Fermi liquids).
The tunneling between leads  and dots is described by the perturbation
$V=V_1+V_2$,
where
\begin{eqnarray}\label{perturbation}
&&V_n=\Gamma\sum_s\left[D^{\dag}_{n,s}c_{n,s}+c^{\dag}_{n,s}D_{n,s}\right],
\nonumber\\
&&D_{n,s}=e^{\pm i\varphi/4}d_{1,s}+e^{\mp i\varphi/4}d_{2,s},
\,\,\,\,\, n=1,2\, ,
\end{eqnarray}
and where the operators $c_{n,s}$ and $d_{n,s}$ annihilate electrons with
spin $s$ in the $n$th lead and in the $n$th dot, {\it resp}.
The Peierls phase $\varphi$ in the hopping
amplitude accounts for an AB or Berry phase (see below) in the
presence of a magnetic field. The upper
sign belongs to lead 1 and the lower  to
lead 2. 
Finally we assume that spin is conserved in the tunneling process.
For the outgoing currents we have
$I_n=ie\Gamma\sum_s\left[D^{\dag}_{n,s}c_{n,s}-c^{\dag}_{n,s}D_{n,s}\right]$.
The observables of interest are the average current
through the DD system, $I=\langle I_2\rangle$,
and the symmetrized cross correlations (noise) of the outgoing currents,
$S(t)=Re\langle\delta I_2(t)\delta I_1(0)\rangle$,
where $\delta I_2=I_2-I$ etc.
First, we evaluate the transport current
and then the noise.

{\it  Cotunneling current.\/}
From now on we concentrate on the CB regime where we can
neglect double (or more)  occupancy in each dot for all  transitions
including virtual ones, {\it i.e.} we require $\mu_{1,2}<U$.
Further we assume that  $\mu_{1,2} >J, k_BT $ to avoid
resonances  which might change the DD state (see
also below).
The lead-dot coupling $\Gamma$   is assumed to be weak so that
the  state of the DD is not perturbed; this will allow us
to retain only the first non-vanishing contribution in $\Gamma$ 
to $I$ and $S(t)$.
Formally, we require  $J> 2\pi\nu_t \Gamma^2$,
where
$\nu_t$ is the tunneling density of states of the leads (see below).

In analogy to the single-dot case\cite{averinazarov,Glattli}  we refer to
above CB regime as  cotunneling regime. In this regime electrons tunnel
one by one
through virtual states of the DD (see Fig.1).
For
real quantum dots not all of our assumptions  might be perfectly satisfied
and phase coherence might be supressed. Still,
since it will turn out that in the cotunneling regime the current has a
phase-coherent part with AB oscillations, it should be possible to
extract this part even if its
amplitude is much smaller than the incoherent one (this is a common situation
in
mesoscopic transport experiments)\cite{YacobyNote}.
Thus, from now on we will
concentrate  on cotunneling only.
We have specified now all assumptions \cite{equilibrium} under which
the following results are valid,
again they are $U>\mu_{1,2}>k_BT,J> 2\pi \nu_t \Gamma^2$. 

Continuing with our derivation of $I$,  we note that the average
$\langle\ldots\rangle\equiv {\rm T}{\rm r} {\rho}\left\{\ldots\right\}$
is
taken with respect to the  equilibrium
state of the {\it entire} system set up in the distant past 
before $V$ is switched
on\cite{Mahan}. Then, in the interaction picture,
the current is given by
\begin{equation}\label{current1}
I=\langle U^{\dag}I_2(t)U\rangle ,\quad U=
{T}\exp\left[-i\int^t_{-\infty} dt' V(t') \right].
\end{equation}
The leading contribution in $\Gamma$ to the cotunneling current
involves the tunneling of
one  electron from the DD to, say, lead
2 and of a second electron from lead 1 to the DD (see Fig.1). This
contribution is of order  $V_2V_1^2$, and thus $I\propto \Gamma^4$, as
is typical for cotunneling\cite{averinazarov}.

After some manipulations we can express the current (\ref{current1}) in
terms of spinless Green functions of the leads evaluated at
coinciding points,
$G^{>}_i(t)=-i\langle c_i(t)c_i^{\dag}(0)\rangle$ and $G^{<}_i(t)=
i\langle c_i^{\dag}(0)c_i(t)\rangle$\cite{Mahan}.
The expanded expression for $I$ is too lengthy to be presented
here, so we
only mention that all terms of the form
$G^{>}_1G^{>}_2$ contain divergencies at small energy transfer around the
DD levels due to resonances\cite{SukhorukovLoss}.
This is seen from the structure of
$G^{>}_i$: first  an electron is created in the leads and then
annihilated, thus describing a resonant current {\it away from} the
DD into leads 1 and 2. If the state in the DD is
maintained
through the coupling to feeding leads $3$ and $4$\cite{SukhorukovLoss},
these resonances will be smeared and the divergencies have to be cut at an
energy
$\varepsilon\sim 2\pi\nu_t{\tilde {\Gamma}}^2$. On the other hand,
in the  cotunneling regime defined above we are
far away from such resonances, and
such a resonant current is suppressed, and
all divergent terms of the form $G^{>}_1G^{>}_2$ vanish.
We arrive then at the following compact expression for the
cotunneling current
\begin{eqnarray}\label{current2}
&&I={e\Gamma^4\over {2\pi}}\sum_{i,f,s,s'}\rho_i\,
|\langle i|D_{2,s'}^{\dag}D_{1,s}|f\rangle|^2
\left[F_{12}-F_{21}\right],  \\
&&F_{12}=\int{{d\varepsilon}\over {\varepsilon^2}}G^{<}_1(\varepsilon +
E_f)G^{>}_2(\varepsilon +E_i).
\label{Fs}
\end{eqnarray}
For simplicity, we assumed that the upper and lower branches of the
tunneling loop (Fig.1) are identical.
Eqs.\ (\ref{current2}) and (\ref{Fs})
show that 
in the cotunneling regime the initial state $|i\rangle$ 
(with
weight
$\rho_i$ and energy $E_i$) of the
DD  is changed into a final state $|f\rangle$ (energy $E_f$)
by the traversing electron.
However, due to the weak coupling $\Gamma$, the DD will have returned to
its equilibrium state before the next electron passes through it
\cite{equilibrium}.
For small bias, $|\mu_1-\mu_2|<J$, only elastic cotunneling
is allowed\cite{Akera}, {\it i.e.}  $E_i=E_f$. However, this regime is not of 
interest here since singlet and triplet contributions turn out to be identical
and thus indistinguishable\cite{SukhorukovLoss}.
We thus focus on the  opposite regime, $|\mu_1-\mu_2|>J$, where inelastic
cotunneling\cite{inelastic} occurs with singlet and triplet contributions being
different.
In this regime we can neglect the
dynamics generated by $J$ compared to the one generated by the bias ("slow
spins"), and drop the energies $E_i$ and $E_f$ in the Eq.\ (\ref{Fs}).
Finally, using
${\rm 1}=\sum_f|f\rangle\langle f|$
we obtain
\begin{eqnarray}\label{current4}
&&I=e\pi\nu_t^2\Gamma^4C(\varphi ){{\mu_1-\mu_2}\over {\mu_1\mu_2}},\\
&&C(\varphi )=\sum_{s,s'}\left[\langle
d^{\dag}_{1s^{\prime}}d_{1s}d^{\dag}_{1s}d_{1s^{\prime}}\rangle +\cos
\varphi\langle d^{\dag}_{1s^{\prime}}d_{1s}d^{\dag}_{2s}d_{2s^{\prime}}
\rangle \right],\label{factor1}
\end{eqnarray}
where we assumed Fermi liquid leads,
$G_n^{>}(\varepsilon )=-i\pi\nu_tf_F(\mu_n-\varepsilon)$,
and $G_n^{<}(\varepsilon )=i\pi\nu_tf_F(\varepsilon -\mu_n)$,
where the tunneling
density of states in the leads $\nu_t=-{2\over
{\pi}}$Im$\sum_pG^R(\varepsilon ,p)$
is expressed in terms of the retarded  Green function\cite{Mahan},
and $f_F$ is the Fermi function.

Eq.\ (\ref{current4})
shows that the cotunneling
current depends on the properties of the equilibrium state of the DD
through
the coherence factor $C(\varphi )$ given in  (\ref{factor1}).
The first term in $C$ is the contribution
from the topologically trivial tunneling path (phase-incoherent part)
which runs from
lead 1 through, say, dot 1 to lead 2 and back. The second
term (phase-coherent part) in $C$ results from an exchange process of
electron
1 with electron 2 via the leads 1 and 2 such that a closed loop is formed
enclosing an area $A$ (see Fig. 1).
Note that for singlet and triplets the initial and final spin states
are the same  after such an exchange process. Thus, in the presence of a
magnetic
field $B$, an AB phase factor  $\varphi=ABe/h$ is acquired.

Next, we evaluate $C(\varphi)$ explicitly
in the singlet-triplet basis:
$|S\rangle ={1\over {\sqrt 2}}\left(|\uparrow\downarrow\rangle -|\downarrow
\uparrow\rangle\right)$, $|T_0\rangle ={1\over {\sqrt 2}}\left(|\uparrow
\downarrow\rangle +|\downarrow\uparrow\rangle\right)$, $|T_{+}\rangle
=|\uparrow\uparrow\rangle$, and
$|T_{-}\rangle =|\downarrow\downarrow\rangle$.
Note that only the singlet
$|S\rangle$ and the triplet $|T_0\rangle$ are entangled EPR pairs while
the remaining triplets are not (they factorize).
Assuming that the DD is in one of these states we obtain the
important
result
\begin{equation}\label{factor2}
C(\varphi )= 2\mp\cos \varphi \,\, .
\end{equation}
Thus, we see that the singlet (upper sign) and the triplets (lower sign)
contribute with {\it opposite sign to the phase-coherent part of the
current}.
One has to distinguish, however,
carefully the entangled from the non-entangled states.
The   phase-coherent part of the entangled states is
a genuine {\it two-particle} effect, while the one of the product states
cannot be distinguished from a phase-coherent {\it single-particle}
effect\cite{YacobyNote}.
Indeed, this follows from the observation that the
phase-coherent part in $C$ factorizes for the product states
$T_{\pm}$ while it does not so for $S, T_0$. Also, for states such as
$|\uparrow\downarrow\rangle$ the coherent part of $C$ vanishes, showing that
two
different (and fixed) spin states cannot lead to a phase-coherent
contribution since we {\it know} which electron goes which part of the loop.

Finally we note that due to the AB phase the role of the singlet and
triplets
can be interchanged which is to say that we can continually transmutate the
statistics of the entangled pairs $S,T_0$ from  fermionic to bosonic
(like in anyons): the symmetric orbital
wave function of the singlet $S$ goes into an antisymmetric one at half a
flux
quantum, and vice versa for the triplet $T_0$.

Next, we allow for equal
population of the singlet and triplet states in the DD,
$\rho_i=1/4$, $i=1,\ldots,4$.
Experimentally,
this can  be achieved e.g. by raising
the tunnel barrier  between the dots
(or by increasing the magnetic field)\cite{Loss98,Burkard}
such that
$J$ vanishes or just becomes smaller than
$k_BT$. Tracing then over the 4 terms we see that the EPR pairs, $S, T_0$,
cancel each other while the unentangled terms, $T_{\pm}$, add up.
[This tracing can also be performed over the standard product spin basis as
typically used for Fermi liquids.]
Thus, in this case we return to the usual Fermi sea situation where we can
no longer
distinguish  single- from two-particle phase-coherence
\cite{buettiker}. We note that this effect can be exploited to search for
entanglement: first prepare the DD in its
entangled ground state (singlet) with $|\mu_1\pm \mu_2|>J>k_BT$,
and then reduce
$J$ below $k_BT$; after some spin relaxation time an equal
population of singlet and
triplet states (all contributing with weight 1/4) is reached with a
concommitant
increase of the current by a factor of 5/2.

We would like to stress that
the amplitude of the AB oscillations is a direct measure of the phase
coherence of the entanglement, while the period via the enclosed area
$A=h/eB_0$
gives a direct
measure  of the non-locality of the EPR pairs,
with $B_0$ being the field at which $\varphi=1$. The triplets themselves
can  be further distinguished by applying a directionally inhomogeneous
magnetic
field (around the loop) producing a Berry phase $\Phi^B$\cite{LossBerry},
which is
positive (negative) for the triplet $m=1 (-1)$, while  it vanishes for the
EPR
pairs $S, T_0$. Thus, for $J=0$ we will eventually see beating in the AB
oscillations due to the positive (negative) shift of the AB phase $\Phi$ by
the
Berry phase,
$\varphi=\Phi \pm \Phi^B$.  We finally note that the closed loop shown in
Fig.1
can actually be made as large as the dephasing length by replacing the
dotted
lines outside the dots e.g. by  wave guides forming a loop with a
lead attached somewhere to it\cite{SukhorukovLoss}.

{\it Shot noise.\/}
We evaluate now  the cross correlations $S(t)$ 
of the  currents by expanding in powers of $V$, see
(\ref{current1}). The first non-vanishing contribution is of order $V_1V_2$.
Under the same assumptions as before,
we calculate the spectral density of the noise, $S(\omega)=\int dte^{i\omega
t}S(t)$. For the zero-frequency noise we obtain
\begin{equation}
S(0)=-{e^2\Gamma^4\over {2\pi}}\!\!\!\sum_{i,f,s,s'}\rho_i\,
|\langle i|D_{2,s'}^{\dag}D_{1,s}|f\rangle|^2 \left[F_{12}+F_{21}\right].
\label{zero}
\end{equation}
We compare now to Eq.\ (\ref{current2}) and note
that depending on the sign of $\mu_1-\mu_2$
either $F_{12}$ or $F_{21}$ vanish (at $T=0$).
Therefore, in the cotunneling regime
the  noise assumes its Poissonian value,
$S(0)=-e|I|$, and we see that the Fano factor (noise-to-current ratio)
is universal and the current and its cross-correlations contain the
same information. [When we allow for
resonances of the type discussed above, this is no longer
the case\cite{SukhorukovLoss}.]

 For finite frequencies in the regime $|\mu_1-\mu_2|>J$ we obtain after
lengthy calculations,
\begin{eqnarray}\label{noise2}
& & S(\omega)=(e\pi\nu_t\Gamma^2)^2
\left[X_{\omega}+X^{*}_{-\omega}\right],
\nonumber \\
& & ImX_{\omega}={{C(\varphi )}\over {2\omega}}\left[\theta (\mu_1-\omega
)-\theta (\mu_2-\omega )\right], \\
& & ReX_{\omega}={{C(\varphi )}\over {2\pi\omega}}sign(\mu_1-\mu_2+\omega
)\ln|{{(\mu_1+\omega )(\mu_2-\omega )}\over {\mu_1\mu_2}}|
\nonumber \\
& & -{1\over {2\pi \omega }}\left[\theta (\omega -\mu_1)\ln
|{{\mu_2-\omega }\over {\mu_2}}|+
\theta (\omega -\mu_2)\ln
|{{\mu_1-\omega}\over {\mu_1}}|\right].
\end{eqnarray}
The real part of $S(\omega)$ is even in $\omega$, while the imaginary part
is odd. A remarkable feature here is that the noise  acquires an imaginary 
(i.e. odd-frequency) part
for finite frequencies,
in contrast to single-barrier junctions, where Im$ S(\omega)$ always
vanishes
since we have $\delta I_1=-\delta I_2$ for all times.
In double-barrier junctions considered here we find that
at small enough bias
$\Delta\mu=\mu_1-\mu_2\ll\mu =\left(\mu_1+\mu_2\right)/2$, the odd part,
Im$ S(\omega)$, given in (\ref{noise2})
exhibits  two narrow peaks at $\omega =\pm\mu$, which in real time lead to
slowly decaying oscillations,
\begin{equation}
S_{odd}(t)=
e^2\pi\nu_t^2\Gamma^4C(\varphi )\frac{\sin(\Delta\mu t/2)}{\mu t}\sin (\mu t).
\end{equation}
These oscillations again depend on the phase-coherence factor $C$ with the
same
properties as discussed before. These oscillations can be interpreted as
a temporary build-up  of a charge-imbalance on
the DD during an uncertainty time $\sim \mu^{-1}$, which results from
cotunneling of electrons and an associated time delay between  out- and
ingoing currents.

Finally, allowing currents to flow
from feeding leads via DD into leads 1, 2 we can arrange for
scattering of unentangled  electrons (as considered previously in
noise \cite{noise}) but now also of entangled ones. In the latter
case we get a non-trivial Fano factor\cite{NoiseEntangle}
due to antibunching (triplets) and bunching (singlet) effects in the
noise\cite{DiVincenzoLossMMM}.

{\it Acknowledgements.} We would like to thank R. Blick, C. Bruder, G.
Burkard,
and D. DiVincenzo for  discussions. This work has been supported
by the Swiss NSF.


\begin{references}
\bibitem{footnoteenergy}
We can express
$E_{s/t}$ in terms of the single-particle energy $\epsilon$ and of
the Coulomb charging energy $E_c=e^2/2C$ (with $C$ being the dot
capacitance),
$E_{s/t}=2\epsilon +2E_C \mp J/2$.
\bibitem{YacobyNote}
We note that  a recent  experiment\protect\cite{Yacoby}
demonstrated that a single quantum dot which is part of an AB-loop
and which operates in the CB regime supports phase-coherent
single-particle transport with AB oscillations.
\bibitem{inelastic}
Note that the AB effect is not suppressed by this inelastic cotunneling,
since the {\it entire} cotunneling process involving also leads is elastic:
the initial and final states of the {\it entire} system have the same energy.
\end{references}
\end{document}